\documentstyle[aps,eqsecnum,preprint,prd,epsf]{revtex}
\def\lq{\left [}
\def\rq{\right ]}
\def\lg{\left \{}
\def\rg{\right \}}
\def\lt{\left (}
\def\rt{\right )}
\def\AA{{\cal A}}
\def\HH{{\cal H}}

\def\overl{\overline}
\newcommand{\be}{\begin{equation}}
\newcommand{\ee}{\end{equation}}
\newcommand{\bea}{\begin{eqnarray}}
\newcommand{\eea}{\end{eqnarray}}
\newcommand{\nn}{\nonumber}

\begin{document}
\addtolength{\jot}{10pt}
\tighten

\draft
\preprint{\vbox{\hbox{BARI-TH/94-174 \hfill}
                \hbox{DSF-T-94/12 \hfill}
                \hbox{INFN-NA-IV-94/12 \hfill}
                 }}
\vskip 1cm
\title{\bf Form factor $A_0(q^2)$,  nonleptonic  $D (B) \to P V$ \\
transitions and  rare $B \to K^*\gamma$ decays \\}
\vskip 0.5cm
\author{Pietro~Colangelo $^{1}$, Fulvia~De~Fazio $^{1,2}$
and Pietro Santorelli $^{3}$\\ }

\vskip 0.5cm

\address{$^{1}$ Istituto Nazionale di Fisica Nucleare, \\
 Sezione di Bari, Italy  \\
$^{2}$ Dipartimento di Fisica dell'Universit\`a di Bari,   \\
via G.Amendola 173, 70126 Bari, Italy \\
$^{3}$ Dipartimento di Fisica dell'Universit\`a ``Federico II'' di Napoli \\
and Istituto Nazionale di Fisica Nucleare, Sezione di Napoli,\\
Mostra D'Oltremare, Pad 19-20, 80125 Napoli, Italy}

\date{August 25, 1994}

\maketitle
\vskip 0.5cm
\begin{abstract}
We use three-point function QCD sum rules to calculate the form factor
$A_0(q^2)$ appearing in the matrix element of the  flavour-changing axial
current between the $D (B)$ state and a vector meson state. We describe
the role of this form factor in nonleptonic $D (B) \to P V$ decays and
analyze the light $SU(3)_F$ symmetry breaking effects. We also
discuss a proposal to relate the branching ratio of  $B \to K^* \gamma$
 to the spectrum of the semileptonic $B \to \rho \ell \nu$ decay.
\end{abstract}

\vspace{1truecm}


\clearpage

\section{Introduction}

The theoretical description of weak exclusive nonleptonic
decays of heavy mesons is carried out, in the factorization
approximation, in three steps. First, an effective
hamiltonian is constructed taking into account the effects of hard gluon
exchanges \cite{altarelli}. Then, the hadronic matrix
elements are factorized into the product of current-particle matrix elements
that can be either inferred from experiment or calculated theoretically
(notice that we only consider two body decays)
\cite{schwinger,bsw1}. Finally, strong rescattering
effects (mainly for $D$ meson decays) are included considering the coupling
to intermediate resonances \cite{buccella}
or using the measured phase shifts \cite{noi1}.

The second step is dictated by our inability in reliably
computing matrix elements of
the effective hamiltonian between the external states involved in the
decay. Let us consider, for example,
the transition $D^+(p) \to {\overl K^{*0}}(p^\prime) \pi^+(q)$,
which is governed by the
effective hamiltonian:
\be
{\cal H}_W = - {G_F \over \sqrt 2}
V^*_{u d} V_{c s} \lq {c_+ + c_- \over 2}  \; O_1 +
{c_+ - c_- \over 2}  \; O_2 \rq
\label{ham} \ee
\noindent
where $G_F$ is the Fermi constant,  $V_{hk}$ are CKM matrix
elements and the local operators $O_1$ and $O_2$ are given by:
\bea
O_1 & = & \bar s \gamma_\mu ( 1- \gamma_5) c \;
\bar u \gamma^\mu ( 1- \gamma_5) d \label{O1} \\
O_2 & = & \bar s \gamma_\mu ( 1- \gamma_5) d \;
\bar u \gamma^\mu ( 1- \gamma_5) c \; . \label{O2} \eea
\noindent The Wilson coefficients
$c_+(\mu)$ and $c_-(\mu)$ account for the hard gluon effects in the
renormalization of $\HH_W$ from $m_W$ to the low energy scale $\mu$;
they are known at the next-to-leading order approximation \cite{buras}.
The amplitude
$\AA (D^+ \to \overl{K}^{*0} \pi^+)$ is written, in the factorization
approximation, as follows:
\bea
\AA (D^+ \to \overl{K}^{*0} \pi^+)_{fact}
 & = & - {G_F \over \sqrt 2} V^*_{ud} V_{cs}
 <\overl{K}^{*0}\pi^+ | \lq c_1 O_1 + c_2 O_2 \rq | D^+>_{fact}  \nn \\
& \equiv & - {G_F \over \sqrt 2} V^*_{ud} V_{cs}
 \Big \{
\lt c_1 + {c_2 \over N_c} \rt
  <\overl {K}^{*0} | \bar s \gamma_\mu (1- \gamma_5) c | D^+>
<\pi^+ |\bar u \gamma^\mu (1- \gamma_5) d |0> \nn \\
& + & \lt c_2 + {c_1 \over N_c} \rt
 <\pi^+ |\bar u \gamma^\mu (1- \gamma_5) c |D^+>
<\overl{K}^{*0} | \bar s \gamma_\mu (1- \gamma_5) d | 0> \Big \}
\label{amp} \eea
\noindent where $c_1 = (c_+ + c_- )/ 2$,
$c_2 = (c_+ - c_-) / 2$. To obtain (\ref{amp}) we have used the
identities
\bea
O_1 & = & \frac{1}{N_c} O_2 +\frac{1}{2}
         \lq \bar s \gamma_\mu (1- \gamma_5)\lambda^a d \rq
         \lq \bar u \gamma^\mu (1- \gamma_5)\lambda^a c \rq \nn\\
O_2 & = & \frac{1}{N_c} O_1 +\frac{1}{2}
         \lq \bar s \gamma_\mu (1- \gamma_5)\lambda^a c \rq
         \lq \bar u \gamma^\mu (1- \gamma_5)\lambda^a d \rq \;
\eea
where $N_c$ is the number of colours and $\lambda^a$ are
the Gell-Mann $SU(3)_c$
matrices. Therefore,
the amplitude is expressed in terms of current-particle matrix elements
\be
<{\overl K}^{*0}(p^\prime, \lambda) | \bar s \gamma_\mu  d | 0>   =
 f_{K^*} m_{K^*}  \epsilon^*_\mu(\lambda) \label{fkstar}
\ee
\be
<\pi^+ (q) | \bar u \gamma_\mu \gamma_5 d | 0>   =
- i f_\pi q_\mu
\ee
\noindent ($\epsilon(\lambda)$ is the $K^*$ polarization vector)
and of the matrix elements governing the semileptonic transition
$D \to K^* (\pi) \ell \nu$:
\bea
& & <\overl{K}^{*0}(p^\prime, \lambda) | \bar s \gamma_\mu (1- \gamma_5) c
| D^+(p)>  =
\Big \{ \epsilon_{\mu \nu \rho \sigma}
\epsilon^{* \nu} p^\rho p^{\prime \sigma} { 2 V(q^2) \over m_D + m_{K^*}}
\nn \\
& - & i \Big[ (m_{D} + m_{K^*}) A_1 (q^2) \epsilon_{\mu}^{\ast}
 -  {A_2 (q^2)
\over m_{D} + m_{K^*}} (\epsilon^{\ast}\cdot p) (p + p^\prime)_\mu \nn \\
& - & \;(\epsilon^{\ast}\cdot p) {2 m_{K^*} \over q^2} q_\mu
\big(A_3 (q^2) - A_0 (q^2)\big) \Big]  \Big \} \; , \label{eqi1} \eea
\be
<\pi^+ (q) | \bar u \gamma_\mu c | D^+(p)> =
 (p + q)_\mu F_1 (p^{\prime 2}) +
{ m^2_D - m^2_\pi \over p^{\prime 2} }
[F_0(p^{\prime 2}) - F_1(p^{\prime 2})] p^\prime_\mu  \; \ee
\noindent
$(q = p - p^\prime)$.
Notice that we have used the BSW \cite{bsw} parameterization
of the form factors; in this parameterization
 $A_1$, $A_2$ and $A_3$ are not
independent, but they fulfil the relation:
\be
A_3 (q^2)\;=\;{m_{D}+m_{K^*} \over 2 m_{K^*}} A_1 (q^2) -
{m_{D}-m_{K^*} \over 2 m_{K^*}} A_2 (q^2) \hskip 5pt  \label{eqi2}\ee
\noindent with
\be A_3(0)=A_0(0) \label{eqi3} \ee
\noindent in order to avoid the unphysical singularity at $q^2=0$
in eq.(\ref{eqi1}); for the same reason $F_0(0)=F_1(0)$.

The amplitude of the process we are considering
is therefore reduced to:
\bea
\AA (D^+ \to \overl{K}^{*0} \pi^+)_{fact} = -
{G_F \over \sqrt 2} V^*_{ud} V_{cs}
2 m_{K^*} (\epsilon^* \cdot p )
& \Big \{ & (c_1 + {c_2 \over N_c}) f_\pi A_0(m^2_\pi) \nn \\
& + & (c_2 + {c_1 \over N_c})  f_{K^*} F_1(m^2_{K^*}) \Big \} \; .
\label{amp1} \eea
\noindent The leptonic constants $f_\pi$ and $f_{K^*}$ can be derived
 from experiment. As for the
Wilson coefficients, it has been observed \cite{bsw1,buras1} that
consistently neglecting all $1/N_c$ terms in the factorized amplitudes like
eq.(\ref{amp1}), and using the form factors e.g. from the BSW model
\cite{bsw},  the overall agreement with the available experimental data seems
to improve. This is an appealing feature of this approach, since
factorization becomes exact in the multicolor
chromodynamics, in the limit $N_c \to
\infty$ \cite{thooft}.
Dynamical justifications of the rule of discarding $1/ N_c$
contributions have been investigated by
analyzing the sign and the size of nonfactorizable matrix elements
in some $D$ and $B$ meson nonleptonic decays\cite{block,kod}.\\

However, in order to have a quantitative overview of the validity of the
factorization approach we need a careful determination, either from experiment
or by a QCD calculation, of the form factors appearing in the factorized
amplitudes like (\ref{amp1}).
This is an important preliminary study in the analysis of the decay
channels where naive factorization seems to fail \cite{cleo}.

Early estimates of $F_1$, $V$, $A_1$ and $A_2$,
evaluated at $q^2=0$ or at $q^2=q^2_{max}$,  are available from constituent
quark models \cite{bsw,gs,is}. More recently, these form factors have been
computed, at a fixed value of $q^2$, by
QCD sum rules \cite{braun,ball,belyaev} and by lattice QCD \cite{reticolo};
 they have also been derived in the framework of models that incorporate
 heavy quark and chiral symmetries \cite{casal}.
Moreover, QCD sum rules have been used to determine
the $q^2$ dependence of
$F_1$, $V$, $A_1$, $A_2$ and $F_0$
\cite{braun,ball,belyaev,santorelli1,marbella}; data on this functional
dependence are also available from lattice QCD, but  with large
error bars \cite{reticolo}.

As for  $A_0(q^2)$, the procedure adopted so far consists in
using  eqs.(\ref{eqi2}-\ref{eqi3}) to get $A_0(0)$ from $A_1(0)$ and $A_2(0)$,
 and then in assuming a suitable dependence
on $q^2$, invoking the dominance of the nearest pole in the $t=q^2$ channel.

This procedure has two difficulties. The first one is that
$A_1(0)$ and $A_2(0)$ are predicted by QCD sum rules and lattice QCD
in a range of values; the uncertainty (larger for $A_2$) is determined by the
variation of the parameters employed in the calculation
(which are not accurately known) and by the statistical error in
lattice simulations. This implies that $A_0(0)$
is determined with a large error (within an order of magnitude for
the transition $B \to \rho$), an uncertainty that heavily
affects the analysis of
the accuracy of the factorization approach. This error is even
more important in the study of light $SU(3)_F$ breaking effects,
an argument which
has recently prompted a number of interesting investigations
\cite{chau}.

The second point concerns the functional dependence of $A_0$ on $q^2$, which
is needed in computing nonleptonic transitions. The assumption that, also
for low values of $q^2$, the form factor $A_0$ is dominated by the nearest
resonance requires an explicit check, since it is known that, for the
transitions $B, D \to \rho, K^*$, the form factors
$A_1$ and $A_2$ obtained by  QCD sum rules appear to be nearly independent of
$q^2$ up to rather large values of the squared momentum transferred
\cite{braun,ball,marbella}.

Both these difficulties can be avoided  by a direct
calculation of $A_0(q^2)$. This is the problem we address in the present paper:
we compute $A_0(q^2)$ for the transitions
$B \to \rho$, $D \to \rho, K^*$ and $B \to D^*$
using three-point function QCD sum rules.
In Sect.2 we derive the sum rule for $A_0(q^2)$, and in Sect.3 we collect
our numerical results, together with a comparison with the outcome of
other calculations.

The last point we analyze in this note is a proposal,
put forward in ref.\cite{BD,oD1,oD2}, to relate the branching ratio of the rare
transition $B \to K^* \gamma$ to the spectrum of the semileptonic decay
$B \to \rho \ell \nu$. This connection could have interesting phenomenological
consequences, e.g. for the measurement of $V_{ub}$ ; it
is based on the equality, obtained in the infinite heavy quark mass limit,
$T_1(0)=  A_0^{B \to \rho}(0)$, where $T_1$ is the form factor governing
$B \to K^* \gamma$. We shall discuss the validity of this relation
and show that there is a sizeable deviation to be taken into account.

\clearpage
\section{The form factor $A_0$ from three point function QCD sum rules.}

In order to calculate the form factor $A_0(q^2)$ in eq.(\ref{eqi1}) using
three-point sum rules,
let us consider the correlator \cite{shiflibro,ioffe}
\be
T^{D \to K^*}_{\mu\nu} (p, p^\prime, q) = i^2 \int dx \; dy \;
e^{i ( p^\prime \cdot x - p \cdot y)} \;
<0| T \{ j_\nu(x) A_\mu(0) j^\dagger_5(y) \} |0> \; . \label{corr}
\ee

Here the weak current $A_\mu =\bar s \gamma_\mu \gamma_5 c$
represents the flavour changing axial current in
(\ref{eqi1}), while the quark currents $j_\nu = \bar d \gamma_\nu  s$
and $j_5 = \bar d i \gamma_5 c$ interpolate  $K^*$ and $D$ mesons,
respectively, and
have a non-vanishing matrix element between the vacuum and $K^*$ and $D$
states: eq.(\ref{fkstar}) and
\be <0|j_5|D(p)>  =  f_D {m_D^2 \over m_c } \; , \ee
\noindent
where $m_c$ is the mass of the charm quark (hereafter we put the masses
 of the light $u$ and $d$ quarks to zero).
The product $q^\mu T_{\mu \nu} $ can be decomposed in two independent
Lorentz structures:
\be
q^\mu \; T_{\mu \nu} = -i \; \Big [ P_\nu \; T_1(p^2, p^{\prime 2}, q^2)
 + q_\nu \; T_2(p^2, p^{\prime 2}, q^2) \; \Big ] \; ,
\ee
\noindent with $P_\nu= (p + p')_\nu$,
and the two scalar functions $T_i$ ($i=1,2$) can be represented by a double
dispersion relation:
\be T_i (p^2, p^{\prime 2}, q^2) =
{1 \over (2 \pi)^2} \int ds \; ds^\prime
{\rho_i^{(had)}(s, s^\prime, q^2) \over (s - p^2) (s^\prime - p^{\prime 2})}
\; + \; subtractions \label{dr1}\ee
\noindent with the spectral functions $\rho_i^{(had)}$ expressed in terms of
physical states:
\be
\rho_i^{(had)} = \rho_i^{(res)}  +  \rho_i^{(hr)}    \; . \label{spectr}
\ee
\noindent
In (\ref{spectr}) $\rho_i^{(res)}$ represent the contribution of the lowest
lying resonances (in the zero width approximation)
and are expressed in terms of the form factor
$A_0^{D \to K^*}$ we are
interested in:
\be  \rho_1^{(res)} (s, s^\prime, q^2) =
(2 \pi)^2 f_{K^*}  f_D {m^2_D \over 2 m_c}
(m^2_{K^*} - m^2_D + q^2) A_0^{D \to K^*}(q^2)
\delta(s-m^2_D) \delta(s^\prime - m^2_{K^*})
\label{rho1} \ee
\be  \rho_2^{(res)} (s, s^\prime, q^2) =
(2 \pi)^2 f_{K^*}  f_D {m^2_D \over 2 m_c}
(3 m^2_{K^*} + m^2_D - q^2) A_0^{D \to K^*}(q^2)
\delta(s-m^2_D) \delta(s^\prime - m^2_{K^*}) \; , \label{rho2}\ee
\noindent
while $\rho_i^{(hr)}$ take contribution from higher states and from
the hadronic continuum.

The correlator (\ref{corr}) can also be calculated in QCD by the operator
product expansion (OPE) in the region of large Euclidean momenta
$p^2$, $p^{\prime 2}$. In this expansion the most singular term is represented
by the perturbative term, which can be obtained evaluating a
triangle diagram.
It has a dispersive representation similar to (\ref{dr1}):
\be T_i^{(pert)} (p^2, p^{\prime 2}, q^2) =
{1 \over (2 \pi)^2} \int ds \; ds^\prime
{\rho_i^{(pert)}(s, s^\prime, q^2) \over (s - p^2) (s^\prime - p^{\prime 2})}
\; ; \label{pert} \ee
\noindent the explicit form of the spectral functions $\rho_i^{(pert)}$ can be
found in the appendix. The next terms in the
OPE can be read as:
\be
T_i^{(np)} = \sum_n d_i^{(n)} (p^2, p^{\prime 2}, q^2) <0| O_n |0> \;
 \label{ope} \ee
\noindent where the local operators $O_n$, written in terms of
quark and gluon fields, are ordered according to their increasing dimension;
their vacuum matrix elements (condensates) parametrize the
effects of the nonperturbative QCD vacuum.
Here we only consider the contribution
of the operators of dimension $D=3$ and $D=5$:
\be
T_i^{(np)} = d_i^{(3)} <\bar q q>  + d_i^{(5)} <\bar q g_s \sigma G q> \; ;
\ee
\noindent
$<\bar q q>$ and $<\bar q g_s \sigma G q>$ are the quark and the mixed
quark-gluon condensates, respectively; the coefficients
$ d_i^{(3)}$ and $ d_i^{(5)}$ can be calculated in perturbation theory.

The hadronic and the QCD representations of the correlator (\ref{corr}) can be
used to derive two sum rules for $A_0(q^2)$. As a matter of fact, we invoke
quark-hadron duality and assume the equality:
\be
\int_{D'} ds \; ds^\prime
{\rho_i^{(hr)}(s, s^\prime, q^2) \over (s - p^2) (s^\prime - p^{\prime 2})} =
\int_{D'} ds \; ds^\prime
{\rho_i^{(pert)}(s, s^\prime, q^2) \over (s - p^2) (s^\prime - p^{\prime 2})}
\; \label{equ} \ee
\noindent
in a region $D'$ above some threshold $s_0$, $s'_0$.
This  assumption can be
realized if we adopt the model:
\be
\rho_i^{(hr)}(s, s^\prime, q^2) =
\rho_i^{(pert)}(s, s^\prime, q^2) ( 1 - \Theta(s_0 - s) \Theta(s'_0 - s') )
\; . \label{mod}
\ee
\noindent
Using (\ref{mod}) the following
sum rules for $A_0(q^2)$ can be written:
\bea
{1 \over (2 \pi)^2} \int_{D} ds \; ds^\prime
{\rho_i^{(res)}(s, s^\prime, q^2) \over (s - p^2)
 (s^\prime - p^{\prime 2})} & = &
{1 \over (2 \pi)^2} \int_{D} ds \; ds^\prime
{\rho_i^{(pert)}(s, s^\prime, q^2) \over (s - p^2)
(s^\prime - p^{\prime 2})} \nn \\
& + & d_i^{(3)} <\bar q q>  + d_i^{(5)} <\bar q g_s \sigma G q> \;
  \label{sr1} \eea
\noindent
where the region $D$ is fixed by:
\be m_c^2 \le s  \le s_0 \label{int1}\ee
\noindent and
\bea
m_s^2 \le & s' & \le s'_0 \nn \\
s^\prime_-(s, q^2) \le & s' & \le s^\prime_+(s, q^2) \label{int2}  \eea
\noindent with ($m_s$ is the strange quark mass):
\bea
s'_{\pm}(s, q^2) & = & \frac{1}{2}\lg\lt s-m_c^2\rt\lt
1-\frac{q^2}{m_c^2}\rt
+\lt\frac{m_s}{m_c}\rt^2\lt s+m_c^2\rt \right.\nn\\
&\pm &\left. \lt s-m_c^2\rt \lq \lt\frac{m_s}{m_c}\rt^4\,+\,
\lt 1-\frac{q^2}{m_c^2}\rt^2 -2 \lt \frac{2m_s^2}{m_c^2} \rt
\lt 1+\frac{q^2}{m_c^2}\rt
\rq^{1/2}\rg \; .
\eea
In (\ref{sr1}) we have omitted the subtraction terms, that can still be
present.
They are removed by performing a double Borel transform to both sides of
(\ref{sr1}) in the variables
$P^2=-p^2$ and $P^{\prime 2}=-p^{\prime 2}$:
\be
B_{P^2, P^{\prime 2}} (M^2, M^{\prime 2}) =
\lq {1 \over n!} (-P^2)^{n+1} \lt {d \over dP^2} \rt^{n+1} \rq
\lq {1 \over m!} (-P^{\prime 2} )^{m+1} \lt {d \over dP^{\prime 2} }
\rt^{m+1} \rq \ee
\noindent in the limit $P^2, P^{\prime 2}  \to \infty$,
$n, m \to \infty$, $M^2 = P^2 /n$ $M^{\prime 2} = P^{\prime 2} /m$ fixed.
The resulting sum rules for $A_0^{D \to K^*}$ read:
\bea
A_0^{D \to K^*} (q^2) & = &
 {H_i \over (2 \pi)^2} \int_D ds \; ds^\prime
\rho_i^{(pert)}(s, s^\prime, q^2) \;
 \exp (- {(s- m^2_D) \over M^2} -
{ (s^\prime - m^2_{K^*})\over M^{\prime 2}} )
 \nonumber \\
 & + & \lq \Gamma_i^{D=3}  \; <\bar q q>  + \Gamma_i^{D=5} \;
<\bar q g_s \sigma^{\mu \nu} G^a_{\mu \nu} {\lambda^a \over 2 } q>
\rq
\; \exp( { m^2_D -m^2_c \over M^2}+{ m^2_{K^*} -m^2_s \over M^{'2}})
\label{sr2} \eea

\noindent ($i=1,2$);
the coefficients $H_i$, $\Gamma_i^{D=3}$ and $\Gamma_i^{D=5}$ are collected in
the appendix.

Eq.(\ref{sr2}) represents two independent sum rules for
$A_0^{D \to K^*} (q^2)$, that will be analyzed in the next section. Now, before
discussing the parameters and the criteria we have used to compute
$A_0^{D \to K^*}$, let us consider the $t$ dependence in (\ref{sr2}).
In principle, the operator product expansion can be reliably applied to
the evaluation of the
correlator (\ref{corr}) in the region af large euclidean values of $q^2$.
However, as discussed in \cite{braun}, we can also consider positive values of
$q^2$ provided that the occurrence of non-Landau
singularities either is avoided or is carefully taken into account.
Such singularities remain far from the integration region of the dispersive
integral if $q^2$
is small with respect to the physical threshold in the $t$ channel:
$t_{th}=(m_c + m_s)^2$; their presence is shown up by
the appearance of branch points (for  large and positive $q^2$) in the spectral
functions $\rho_i^{(pert)}$.

Taking into account the above considerations,  we study
the form factor $A_0$ using eq.(\ref{sr2})  for small positive
values of $q^2$, and not in the whole physically accessible $t$-region.
However, the information we obtain on the $t-$dependence of the form
factor is enough to describe a large number of nonleptonic $D(B)-$meson decays.

\clearpage
\section{Numerical analysis of the sum rules}
The numerical analysis of the sum rules (\ref{sr2}) is performed using
the following values for the quark condensates, taken at a
low renormalization scale ($\mu \simeq 1 \; GeV$) \cite{shiflibro}:
\bea
<\bar q q> & = & (- 230 \; MeV)^3 \nn \\
<\bar q g_s \sigma^{\mu \nu} G^a_{\mu \nu} {\lambda^a \over 2 } q> & = &
m_0^2 \; <\bar q q> \label{cond}\eea
\noindent with $m_0^2=0.8 \; GeV^2$. The condensates can be
 evaluated at higher
scales using the leading-log approximation for the anomalous dimension
of the quark and of the mixed quark-gluon condensates; however,
this rescaling does not affect sensitively the numerical result for
$A_0$.

As for the quark masses, we use:
$m_c=1.35 \; GeV$ and  $m_s=0.16 \; GeV$; moreover, we use
$ m_b=4.6  \; GeV$ in the calculation of form factor connected with
the transition $B \to \rho$.
The leptonic constant $f_{K^*}$ can be derived from the measurement of the
branching ratio $\tau \to K^* \nu_\tau$:
$f_{K^*}~=~0.22\pm0.01~GeV$;
we use $f_{D} = 195 \pm 20  \; MeV$
for the leptonic constant of the $D$ meson.

A comment on this value of $f_D$ is in order. This value
comes from two-point QCD sum
rules, including radiative corrections at the order ${\cal O}(\alpha_s)$
\cite{colangelo1,dominguez}; these corrections are at level of $13-15\;
\%$ for $f_D, f_{D^*}$ and play a major role in $f_B$; including the
corrections the results are in agreement with lattice QCD. On the other hand,
the QCD expression of the three-point correlator has been computed at zero
order in $\alpha_s$. Several authors \cite{braun}, when
 computing the form factors,
adopt the strategy of taking the ratio of three and two-point functions,
calculated
at the same order in $\alpha_s$ in order to be consistent, trying to reduce the
uncertainty deriving from higher order contributions. However, the
sign of the
radiative corrections to three point functions is not known on general
grounds; for this reason our attitude is to use the best known values of the
leptonic constants, e.g. the values on which QCD sum rules and lattice QCD
agree.

The effective threshold $s_0$ and $s_0^\prime$ must be chosen in a range of
values between the mass squared of the lowest lying resonances and the first
excited states.  They have been fixed by studying two-point sum rules,
for the calculation of static properties as the mass of the leptonic constant
of the particles,
and three-point functions sum rules in the calculation of the semileptonic
$D \to K^* \ell \nu$ decay \cite{braun}: $s_0 = 7 - 8 \; GeV^2$ for the channel
of $D$ meson, and $s_0^\prime= 1.5 - 1.7 \; GeV^2$ for the channel of $K^*$.
The variation of the values of $s_0, s_0^\prime$ induces a variation of the
predicted form factor $A_0$ that can be considered a theoretical
uncertainty in the final result.

Using the parameters above, $A_0^{D \to K^*} (q^2)$ can be obtained from
(\ref{sr2}) as a function of the Borel parameters $M^2$ and $M^{\prime 2}$.
However, since these variables are unphysical, we search a region in
$M^2, M^{\prime 2}$ where $A_0$ does not depend on them (stability
plateau). In this region other conditions must be verified. The
first constraint consists in checking a hierarchical structure of the various
terms of (\ref{sr2}):
\be
T^{(pert)} \ge T^{(D=3)} \ge T^{(D=5)} \; ;\label{cond1} \ee
\noindent only if the OPE displays this structure
we can hope that higher order
power corrections in (\ref{sr2}) can be safely neglected.

Another condition is connected with the approximation (\ref{equ}).
Since we are not guaranteed that quark-hadron duality starts already in
correspondence to the first hadronic excitations, we must choose
$M^2, M^{\prime 2}$ small enough to enhance the contribution of the lowest
lying states in the sum rule, and to exponentially suppress the contribution of
the continuum. We  check this condition by verifying that
\be
\left | \int_D ds \; ds^\prime
\rho_i^{(pert)}(s, s^\prime, q^2) \;
 e^{ - {(s- m^2_D) \over M^2} -
{ (s^\prime - m^2_{K^*})\over M^{\prime 2}} }\right |  \ge \alpha \;
\left | \int_{D_1} ds \; ds^\prime
\rho_i^{(pert)}(s, s^\prime, q^2) \;
 e^{ - {(s- m^2_D) \over M^2} -
{ (s^\prime - m^2_{K^*})\over M^{\prime 2}} }\right |  \label{cond2} \ee
\noindent where $D_1$ is an integration region larger than $D$
(it extends to $10 \; s_0$ and $10 \; s^\prime_0$) and $\alpha=0.4-0.5$.
{}From the condition (\ref{cond2}) an upper bound
to $M^2, M^{\prime 2}$ can be determined.

In principle both the sum rules $(i=1,2)$ in (\ref{sr2}) can be used to
calculate $A_0$. However, it turns out that only for the first sum rule
$(i=1)$ the condition (\ref{cond2}) is fulfilled. In the second sum rule
the constraint (\ref{cond2}) is never verified
as a consequence, probably, of the major role played by higher
states. For this reason we exclude the second sum rule in our
analysis and use only the first one to determine $A_0$.

Let us consider the  form factor $A_0^{D \to K^*}(q^2)$.
Our results, obtained for
$M^2=3 \; GeV^2$
and $M^{\prime 2}=1.5 \; GeV^2$, are  depicted in fig.(1) for several sets of
parameters. At $q^2=0$ we have:
$A_0^{D \to K^*}(0)=0.58 \pm 0.05$, where the uncertainty comes from the
variation of the parameters. For positive values of $q^2$ we observe an
increasing  of the form factor; if we use the simple pole formula:
\be
A_0^{D \to K^*}(q^2)= {A_0^{D \to K^*}(0) \over 1 - q^2/m_P^2} \label{pole}
\ee
\noindent
to fit the $q^2$ dependence, we obtain that
 the fitted pole mass is: $m_P= 1.65 \; GeV$, which
is not far from the experimental mass of
the particle having the same quantum numbers of the pole:
$m_{D_s}=1.969 \; GeV$.

Let us now consider the transition $D \to \rho$.
The constant $f_{\rho^+}$, computed from the decay
$\rho^0 \to e^+ e^-$ and assuming the isospin symmetry, is given by:
$f_{\rho^+}=216 \; MeV$. Using $s^\prime_0=1.3 - 1.5 \; GeV^2$ we get:
$A_0^{D \to \rho}(0)=0.52 \pm 0.05$. The fitted pole mass is
$m_P= 1.6 \; GeV$, to be compared with $m_D=1.865 \; GeV$ which is the mass of
the nearest pole in the $t$ channel (fig.(2)).

{}From the above results we would get:
$r=A_0^{D \to K^*}(0)/A_0^{D \to \rho}(0)=1.12 \pm 0.11$;
however, the light $SU(3)_F$ breaking effects can be better
estimated by studying the ratio of the sum rules for the transitions
$D \to K^*$ and $D \to \rho$, which is stable with respect to the
variation of the input parameters. We get $r=1.10 \pm 0.05$, i.e.
an $SU(3)_F$ breaking effect at the level of  $10 \%$.

We have also computed the form factor
$A_0^{B \to \rho}$ appearing in nonleptonic $B$ meson decays which are
interesting due to their connection to the measurement of $V_{ub}$; using
$s_0 = 33-36 \; GeV^2$ and $f_B=180 \pm 20 \; MeV$
as obtained by two point function QCD sum rules, we get
$A_0^{B \to \rho}(0) = 0.24 \pm 0.02$  and a pole mass
$m_P=5 \; GeV$,
to be compared to $m_B=5.275 \; GeV$; the form factor is depicted in
fig.(3) in the range $q^2 = 0 - 10 \; GeV^2$.

Finally, for the heavy-to-heavy meson transition $B \to D^*$ we get
(using $f_{D^*}=250 \; MeV$ \cite{colangelo1}):
$A_0^{B \to D^*}(0)=0.65 \pm 0.05$.

Let us now compare our results with existing  predictions for $A_0(0)$,
obtained from the
values of $A_1(0)$ and $A_2(0)$ in the corresponding channels
(see Table I). The first observation is that our predictions
appear to be smaller than from potential models
\cite{bsw,gs,is} (the exception is $A_0^{B \to \rho}$ in \cite{gs}).
On the other hand, the $SU(3)$ breaking effects are of the same size as in our
approach.

The comparison with QCD sum rules  predictions \cite{braun,ball} shows
the problem we have mentioned in the introduction: the uncertainties
in the results obtained using $A_1(0)$ and $A_2(0)$ are so large that
they obscure the real value of $A_0(0)$,
whereas the error in our predictions is at the level
of $10 \%$, and therefore a substantial improvement has been obtained.

It is interesting to compare our predictions to the outcome of \cite{casal}
obtained in a framework based on heavy-quark and chiral
symmetries, using as an input the experimental data on $D \to K^* \ell \nu$.
Although there is a remarkable agreement on
$D \to K^*$ and $B \to \rho$, the light
 $SU(3)_F$ breaking corrections connecting $D \to K^*$ to  $D \to \rho$
are of similar size than in our approach but with opposite sign.

The last point we would like to mention is that, in our calculation, the form
factors $A_0$ in the different channels increase for positive values of
$q^2$, and that their functional dependence is compatible with the simple pole
behaviour dominated by the nearest resonance in the $t$ channel.

\section{$B\rightarrow K^*\gamma$ versus $B\rightarrow \rho \ell \nu$}
In this section we discuss a procedure, proposed in refs.\cite{BD,oD1},
to relate the width of the radiative $B \to K^* \gamma$ transition to the
spectrum of the Cabibbo suppressed semileptonic decay $B \to \rho \ell \nu$.
This relation can be used to constrain the electroweak parameters involved
in these processes and to reduce their dependence on the models for the
hadronic form factors.

The radiative rare $B$ meson decays, like
$B \to K^*_i \gamma$ ($K^*_i=K^*(890), K_1(1400)$, etc.), have been extensively
studied,  from the theoretical  standpoint, since they have a peculiar role
in the  precision tests of the quark sector in the Standard Model (SM)
and probe the effects of new physics beyond the Standard Model
\cite{review}. Within SM they are induced by
the one-loop electromagnetic penguin operator $b \rightarrow s \gamma$
(for $m_s << m_b$) \cite{1}:
\be
{\cal H}_{eff} (b \to s) = {G_F \over \sqrt 2}
C m_b \epsilon^\mu {\bar s} \sigma_{\mu\nu}
{1 + \gamma_5 \over 2} q^\nu b
\ee
\noindent
where the constant $C$ contains the dependence on the Cabibbo-Kobayashi-Maskawa
matrix elements, on the QCD correcting terms and on the ratio
$m_{top}/m_W$ (the explicit formulae can be found in ref.\cite{1}.

The description of the exclusive decays requires the knowledge of
form factors. For example,  the relevant matrix element for the transition
$B \to K^*(890) \gamma$ can be written as follows:
\clearpage
\bea
\langle K^*(p^\prime,\lambda)|\bar{s} \sigma_{\mu\nu}q^{\nu}
{(1 + \gamma_5) \over 2} b |B(p)\rangle
&=& i\varepsilon_{\mu\nu\rho\sigma}
\epsilon^{*\nu}p^{\rho} p^{\prime \sigma} \; T_{1}(q^{2}) \nonumber\\
&& + \left[(m^{2}_{B}-m^{2}_{K^*})\epsilon^{*}_{\mu}-
(\epsilon^{*}\cdot q)(p+p^\prime)_{\mu}\right]T_{2}(q^{2})\nonumber\\
&& + (\epsilon^{*}\cdot q)\left[(p-p^\prime)_{\mu}-\frac{q^{2}}{(m^{2}_{B}
-m^{2}_{K^*})}(p+p^\prime)_{\mu}\right]T_{3}(q^{2}) \;.
\eea
\noindent
The form factors $T_i(q^2)$ require a nonperturbative evaluation.
However, in the static limit for the $b$ quark and using the $SU(3)_F$ symmetry
for the light quarks, a number of interesting relations can be derived between
$T_i$ and the form factors $V$ and $A_i$ governing $B \to \rho \ell \nu$
\cite{BD,IW1}.
Although such relations are obtained at zero recoil
($q^2_{max}= (m_B - m_{K^*, \rho})^2\;$),
it has been argued  that they continue to
hold in the whole physically accessible $q^2$ range
\cite{BD,ColangeloBK*g,ali}, and therefore that one is able to
relate the width of $B \to K^* \gamma$ to observable quantities in
$B \to \rho \ell \nu$. An example is represented by the ratio:
\be
\Gamma(B\rightarrow K^{*}\gamma)\left( \lim_{q^2\rightarrow 0,curve}
\frac{1}{q^2} \frac{d\Gamma (B \rightarrow \rho e \bar{\nu})}{dE_{\rho}dE_e}
\right) ^{-1}
=\frac{4 \pi^2}{G^{2}_{F}}\frac{|\eta|^2}{|V_{ub}|^{2}}
\frac{(m_{B}^{2}-m^{2}_{K^{*}})^3}{m_{B}^{4}}\,\,\label{eqBD}
\ee
that Burdman and Donoghue \cite{BD} predict to be
independent of any form factor. In (\ref{eqBD}) $curve$
means the region in the Dalitz plot where $q^2=4\,E_e\,(m_B-E_{\rho}-E_e)$;
the factor $\eta$ includes
the $QCD$ corrections to the $b\rightarrow\, s\gamma$
decay and the other relevant electroweak parameters \cite{BD}.

Eq.(\ref{eqBD})  is interesting, both from the experimental and the theoretical
point of view, since it would provide, for example,
a method to measure $V_{ub}$ without referring to calculations of
the form factors, from the
measurement $BR(B \to K^* \gamma)=(4.5 \pm 1.0 \pm 0.9) \times
10^{-5}$ \cite{CLEObsg} and the spectrum of
$B \to \rho \ell \nu$. The drawback, pointed out by O'Donnell and Tung
\cite{oD1}, is that  the semileptonic decay spectrum vanishes at $q^2=0$
on this {\it curve}, with the consequence that the number of experimental
points required to measure the denominator in (\ref{eqBD}) could be
insufficient for a reliable determination of $V_{ub}$.

A more convenient observable \cite{oD2,Santorelli} is represented by
the ratio:
\be
 R(B\rightarrow K^*\gamma) \left(\left.
\frac {d\Gamma(\bar B\rightarrow \rho l\bar\nu_l)}{dq^2}
\right|_{q^2=0}\right)^{-1}   =
 \frac{192\pi^3}{G_F^2} \frac{1}{|V_{bu}|^2}
\frac{(m_B^2-m_{K^*}^2)^3}{(m_B^2-m_{\rho}^2)^3}
\frac{m_b^3}{(m_b^2-m_s^2)^3}|{\cal I}|^2~ \;.
\label{ratio1}
\ee
In (\ref{ratio1})
$R(B\rightarrow K^*\gamma)={\Gamma(B\rightarrow K^*\gamma)}/
{\Gamma(b\rightarrow s\gamma)}$ is
independent of the electroweak parameters appearing
in the effective Hamiltonian,
and only depends on  $T_1(0)$:
\bea
R(B\rightarrow K^{*}\gamma ) =
\frac{m_{b}^{3}(m_{B}^{2}-m_{K^{*}}^{2})^{3}}
{m_{B}^{3}(m_{b}^{2}-m_{s}^{2})^{3}} \; |T_{1}(0)|^2 \; ,
\eea
having used the relation $T_2(0) = \frac{1}{2} T_1(0)$. The factor
${\cal I}$ is given by \cite{oD2}:
\be
{\cal I} = \frac{(m_B+m_{\rho})}{(m_B+m_{K^*})}\,
\frac{T_1(0) }{A^{B\rightarrow\rho}_0(0)}\,\,.
\label{ratio2}
\ee

The main reason for studying the ratio (\ref{ratio1}) is that
${d\Gamma(\bar B\rightarrow \rho l\bar\nu_l)}/{dq^2}$
in (\ref{ratio1}) does not vanish at $q^2=0$.
Moreover, the factor $\cal I$ should be close to one.
As a matter of fact, in the framework of the Bauer-Stech-Wirbel model
\cite{bsw} the value $ {\cal I} = 1.12$ can be derived,
whereas the constituent quark model \cite{oD3} provides
the value $ {\cal I} = 1.09-1.18$.
The relation ${\cal I}=1$ has been obtained by the authors in ref.\cite{oD1}
by considering the transition $b {\bar q} \to Q \bar q$ in the limit where
both the $b$ and $Q$ quarks are heavy, applying spin symmetry relations;
then, it has been argued that this relation
still holds for a light $Q$ quark in the
weak binding limit for the meson.

The factor
$ {\cal I} $ can be derived by three point QCD sum rules, using realistic
values of the quark masses in a fully relativistic approach.
Using the determination $T_1 (0)=0.35 \pm 0.05$ \cite{ColangeloBK*g}
and the value  $A_0^{B\rightarrow\rho} (0)$ derived in the previous section,
we get: $ {\cal I} = 1.43 \pm 0.32 \;$. Notice that,
in evaluating the error for $\cal I$, we have taken into account the
correlated errors in the numerator and in the denominator of $\cal I$.

The uncertainty  can be reduced by studying the ratio of the sum rules
determining $T_1 (0)$ (the formulae can be found in
\cite{ColangeloBK*g}) and $A_0^{B \to \rho}(0)$.
We find a wide stability window in correspondence to the result:
$ {\cal I} = 1.3 \pm 0.1 \;. $

Therefore, the QCD sum rules prediction differs by
$10-30\%$ from the values derived in \cite{oD1,oD2}; this deviation
still is model dependent, in the sense that it comes out from
an explicit QCD sum rules calculation and not from symmetry arguments,
and therefore it should be
checked in a different approach, e.g. by lattice QCD.

In any case, we feel that,
modulo this uncertainty on  the value of $\cal I$,
the idea of relating $B \to K^* \gamma$ to $B \to \rho \ell \nu$
provides us with an interesting alternative method to measure $V_{ub}$.

\section{Conclusions}

We have applied three-point function QCD sum rules to calculate
the form factor $A_0$ appearing in the matrix element of the flavour
changing axial current between
the $D$ and $B$ mesons and a vector meson state.
This form factor plays an important role in the calculation
of non leptonic
$B$ and $D$ decays to $P V$ states in the hypothesis of factorization.

The explicit calculation has allowed us to determine $A_0$
at zero momentum transferred,
with better accuracy than in the indirect
determination from $A_1$ and $A_2$.
We have also computed the $q^2$ dependence, finding
an increase for positive values of the squared
momentum transferred, compatible with a simple polar behaviour.

Finally, we have considered the role of $A_0$ in relating the rare
$B \to K^* \gamma$ decay with the semileptonic $B \to \rho \ell \nu$ spectrum,
in  the search of methods for measuring  the matrix element $V_{ub}$.

\acknowledgments

We thank F.Buccella, G.Nardulli and N.Paver for discussions.

\newpage
\appendix
\section{Perturbative and non-perturbative terms.}

The perturbative spectral functions $\rho_i^{(pert)}(s,s^\prime,q^2)$ in
(\ref{pert}),  at the lowest order in $\alpha_s$,
can be obtained by the Cutkosky rules applied to a
triangle diagram:

\bea
\rho_1(s,s',q^2)& = &\frac{3}{4\sqrt{\lambda}}
\Big \{ (m_c+m_s)\,(\Delta'-\Delta)
+\frac{2(\Delta\,s'+\Delta'\,s)-u\,(\Delta'+\Delta)}{\lambda}
\nn \\
&\times & \lq -(m_c+m_s)(s-s'+q^2)+2m_c(m_c^2-m_s^2)\rq \Big \} \;,
\nn \eea
\bea
\rho_2(s,s',q^2)& = &\frac{3}{4\sqrt{\lambda}}
\Big \{ (m_c+m_s)\,(\Delta'+\Delta)
+\frac{2(\Delta\,s'-\Delta'\,s)+u\,(\Delta-\Delta')}{\lambda}
\nn\\
&\times &\lq (m_c+m_s)(-s+s'-q^2)+2m_c(m_c^2-m_s^2)\rq\Big \} \;,
 \nn \eea
\noindent where
$\Delta = s-m_c^2$, $\Delta' =s'\, -\,  m_s^2$, $u = s + s' - q^2$ and
$\lambda = u^2-4\,s\,s'$.

The Borel transformed coefficients of the non-perturbative
$D=3$ terms read:
\be
\Gamma_1^{(D=3)} =  {m_c^2-m_s^2 \over 2} \nonumber \ee
\be
\Gamma_2^{(D=3)} = - {(m_c+m_s)^2 \over 2} \; ; \nonumber \ee
\noindent the coefficients of the condensate of dimension $D=5$ terms are:
\bea
\Gamma_1^{(D=5)}
& = &\frac{m_c^2 (m_s^2-m_c^2)}{8\,M^4} +
  \frac{m_s^2 (m_s^2-m_c^2)}{8\,M^{'4}} +
  \frac{(m_c^2-m_s^2)}{6\,M^2} +
  \frac{(m_c^2-m_s^2)}{12\,M^{'2}}  \nn \\
&+ & \frac{(m_s^2-m_c^2) \, (2 m_s^2-m_c^2) - q^2 (m_s+m_c) (2 m_s - m_c)}
{12 \, M^2 \, M^{'2}}
\nn \eea
\noindent and
\bea
\Gamma_2^{(D=5)}
& = &\frac{m_c^2 (m_s+m_c)^2}{8\,M^4} +
  \frac{m_s^2 (m_s+m_c)^2}{8\,M^{'4}} -
  \frac{(m_c^2+ 3 m_s^2 +4 m_s m_c)}{12\,M^{'2}} -
  \frac{(3 m_c^2+m_s^2 +4 m_s m_c)}{6\,M^{2}}  \nn \\
&+ & \frac{(m_s^2+m_c^2 -q^2) \, (2 m_s^2+m_c^2) + m_s m_c (4 m_c^2 +2 m_s^2
-3 q^2)}
{12 \, M^2 \, M^{'2}} \; .
\nn \eea

Finally, the coefficients $H_i$ in (\ref{sr2}) read:
\be
H_1 = \frac{2 m_c} {f_{K^*} f_D m_D^2 (m_{K^*}^2 - m_D^2 + q^2)}
\nonumber \ee
\noindent and
\be
H_2 = \frac{2 m_c} {f_{K^*} f_D m_D^2 (3 m_{K^*}^2 + m_D^2 - q^2)} \; .
\nonumber \ee

The formulae for the transitions $D \to \rho$, $B \to \rho$
and $B \to D^*$ can be easily obtained.

\clearpage

\begin{table}
\caption{The form factor $A_0(0)$ of various transitions  in
different models.}
\begin{tabular}{llllllll}
Transition & \cite{bsw} &\cite{gs}& \cite{is}&\cite{ball}&
\cite{reticolo} & \cite{casal} & This paper \\ \tableline
$D \to K^* $ &0.74 &  0.91 & 0.8& $0.45 \pm  0.30$&$ 0.77 \pm 0.29$ & 0.59 &
$0.58 \pm 0.05 $\\
$ D \to \rho $ & 0.68 & -- & 0.85& $ 0.57 \pm 0.40$ & -- & 0.74 &
$0.52 \pm 0.05$\\
$B \to \rho $ & 0.28 & -- & 0.14& $0.79 \pm 0.80$&$ -0.57 \pm 0.65$ & 0.24 &
$ 0.24 \pm 0.02 $ \\
\end{tabular}
\end{table}

\newpage

\hskip 3 cm {\bf FIGURE CAPTIONS}
\vskip 1 cm
{\bf Fig. 1} \par
\noindent
The form factor $A_0^{D \to K^*}(q^2)$.
The curves refer to different sets of parameters:
$s_0=7 \; GeV^2$ and $s^\prime_0=1.5 \; GeV^2$ (continuous line),
$s_0=7 \; GeV^2$ and $s^\prime_0=1.7 \; GeV^2$ (dashed line),
$s_0=8 \; GeV^2$ and $s^\prime_0=1.5 \; GeV^2$ (dotted line),
$s_0=8 \; GeV^2$ and $s^\prime_0=1.7 \; GeV^2$ (dashed-dotted line).
$M^2=3 \; GeV^2$, $M^{\prime 2}=1.5  \; GeV^2$.
\vskip 1 cm
{\bf Fig. 2} \par
\noindent
The form factor $A_0^{D \to \rho}(q^2)$.
The curves refer to the sets of parameters:
$s_0=7 \; GeV^2$ and $s^\prime_0=1.3 \; GeV^2$ (continuous line),
$s_0=7 \; GeV^2$ and $s^\prime_0=1.5 \; GeV^2$ (dashed line),
$s_0=8 \; GeV^2$ and $s^\prime_0=1.3 \; GeV^2$ (dotted line),
$s_0=8 \; GeV^2$ and $s^\prime_0=1.5 \; GeV^2$ (dashed-dotted line).
$M^2=3 \; GeV^2$, $M^{\prime 2}=1.5  \; GeV^2$.
\vskip 1 cm
{\bf Fig. 3} \par
\noindent The form factor $A_0^{B \to \rho}(q^2)$.
The curves refer to the sets of parameters:
$s_0=33 \; GeV^2$ and $s^\prime_0=1.3 \; GeV^2$ (continuous line),
$s_0=33 \; GeV^2$ and $s^\prime_0=1.5 \; GeV^2$ (dashed line),
$s_0=36 \; GeV^2$ and $s^\prime_0=1.3 \; GeV^2$ (dotted line),
$s_0=36 \; GeV^2$ and $s^\prime_0=1.5 \; GeV^2$ (dashed-dotted line).
The values of the Borel parameters are: $M^2=8  \; GeV^2$,
$M^{\prime 2}=2 \; GeV^2$.
\end{document}